\documentclass[twocolumn,amsmath,amssymb,epsfig,prl,aps]{revtex4}

\usepackage{graphicx}
\usepackage{dcolumn}
\usepackage{bm}
\newcommand{\abs}[1]{\left\vert#1\right\vert}

\begin{document}

\title{Parametrically driven dark solitons}
\author{I.V. Barashenkov}
\author{S.R. Woodford}
\author{E.V. Zemlyanaya}
\affiliation{Department of Applied Mathematics,
University of Cape
Town, Rondebosch 7701, South Africa}

\begin{abstract}
We show that unlike the bright solitons, 
the  parametrically
driven kinks are immune from instabilities
for all dampings and forcing amplitudes; they can also form stable
  bound states.
 In the undamped case, the two types of stable kinks 
 and their complexes can travel with nonzero velocities.
 \end{abstract}


\maketitle

The parametric driving is well known to be an efficient way of
compensating dissipative losses of solitons in various media. Examples
include surface solitons in vertically oscillating layers of water
\cite{Faraday,Elphick_Meron};
 light pulses in optical fibers under phase-sensitive
amplification \cite{fibers} and in Kerr-type optical parametric
oscillators \cite{Longhi};
 magnetisation solitons
in easy-plane ferromagnets exposed to oscillatory magnetic fields
in the easy plane
\cite{BBK}.
 A serious problem associated with the
parametric energy pumping, however, is that the  driven solitons
are prone to  oscillatory instabilities which set in as the driver's
strength exceeds a certain --- often rather low --- threshold
\cite{BBK,oscillatory}.

With  a few notable exceptions, the parametrically driven solitons
considered so far had the form of pulses decaying to zero at spatial
infinities. These  were solutions of the nonlinear Schr\"odinger
(NLS)
equation with the ``self-focusing" nonlinearity:
\begin{equation}
i \psi_t + \psi_{xx}  + 2 |\psi|^2 \psi - \psi =
h {\psi^*}- i \gamma\psi,
\label{NLS_attractive}
\end{equation}
where $\psi$ is 
 the amplitude of a nearly-harmonic stationary
wave oscillating with half the driving frequency, 
$\gamma$   the damping coefficient, $h$  
the driving strength, and $^*$ indicates complex conjugation.
However, in a number of applications the amplitude equation
 of the parametrically driven wave turns out to have the
nonlinearity of the  ``defocusing" type:
\begin{equation}
i \psi_t + {\textstyle \frac12} \psi_{xx} - |\psi|^2 \psi + \psi =
h {\psi}^*- i \gamma\psi.
\label{NLS}
\end{equation}
In fluid dynamics, the ``defocusing" parametrically driven NLS
(\ref{NLS}) describes the amplitude of
the oscillation of the water surface in a vibrated channel with a large 
width-to-depth ratio \cite{Elphick_Meron,Larraza}. (On the contrary, 
the ``focusing"
equation (\ref{NLS_attractive}) pertains to the case of narrow
channels.)
The same equation (\ref{NLS}) arises as an
amplitude
equation for the upper cutoff mode in the parametrically driven damped
nonlinear lattices \cite{lattices}.
 In the optical context, it was derived for the
doubly resonant $\chi^{(2)}$
optical parametric oscillator in the limit
of large second-harmonic detuning \cite{Trillo}. 
Next,  in the absence of damping, 
stationary solutions $\psi=M_y+iM_z$ of eq.(\ref{NLS}) minimise the 
Ginzburg-Landau 
free energy for the anisotropic $XY$ model,
$F=\int {\cal F}d {x}$, where
\begin{equation}
{\cal F} = 
{\textstyle \frac12} (\partial_x {\bf M})^2 -(1+h){\bf M}^2 + {\textstyle
\frac12} {\bf M}^4  +
2 h M_y^2 + {\cal F}_0,
\nonumber
\end{equation}
and ${\bf M}=(0,M_y,M_z)$.
This model was used to study domain walls in easy-axis ferromagnets
near the Curie point \cite{XY}. 
 Nonstationary 
magnetisation configurations
were considered in the overdamped limit: $\psi_t=
-\delta F/\delta {\psi}^*$ \cite{Coullet,Coullet_Emilsson}.
The damped hamiltonian dynamics $\psi_t=-i\delta F/\delta {\psi}^*
-  \gamma \psi$ provides a sensible alternative; this is precisely
our eq.(\ref{NLS}).
Finally, for $\gamma=0$ there is yet another, independent,
 magnetic interpretation of eq.(\ref{NLS});
this will be introduced below.

Localised structures characteristic of a defocusing
medium are  domain walls, or kinks, 
also known as ``dark solitons" in the context of nonlinear optics. The
purpose of this note is to explore the stability
and bifurcations of the
parametrically driven kinks and their bound states.

Two stationary kink solutions of (\ref{NLS}) are available
in literature. One is usually called the N\'eel, or Ising, wall:
\begin{equation}
\psi_{N}(x)  = iA \tanh (Ax) e^{-i \theta}.
\label{3} 
\end{equation}
Here
$A^2 = 1 + \sqrt{h^2 - \gamma^2}$ and
 $\theta =  -
 \frac{1}{2} \arcsin \frac{\gamma}{h}$
 \cite{Elphick_Meron,Larraza,Trillo}.
For $\gamma=0$, the N\'eel wall coexists with the Bloch wall:
\begin{equation}
\psi_B(x) = -i A \tanh(2\sqrt{h}x) \pm \sqrt{1 - 3 h} \,
\mbox{sech}(2\sqrt{h}x),
\label{19}
\end{equation}
$A^2=1+h$ \cite{Sarker}.
Originally, the ``magnetic" terminology 
was motivated merely by the fact that $|\psi|=0$ 
in the core
of the wall (\ref{3}), and so the point $x=0$ is a phase defect 
similar to N\'eel points in solid state physics
\cite{Coullet_Emilsson}. On the contrary, in the core of the Bloch
wall the phase changes smoothly --- and this is 
analogous to Bloch walls in ferromagnets. Below we  show that the
analogy with magnetism is in fact much deeper than originally thought.
Letting $h =\gamma = 0$, the N\'eel wall becomes the usual, undriven, 
 dark 
soliton whereas the Bloch wall degenerates into a flat solution. 

Both Bloch and N\'eel walls admit a clear interpretation in other 
physical contexts as well. For example, when eq.(\ref{3}) is 
used to model the Faraday resonance in water \cite{Elphick_Meron,Larraza}
or chains of coupled pendula \cite{lattices}, 
both solutions describe transitions between two domains oscillating
$180^\circ$
out of phase. The phase of the oscillation 
is discontinuous across 
the N\'eel wall and the amplitude  changes over a narrow region;
hence the wall appears as a highly localised defect. 
Conversely, the Bloch wall has a smooth helicoidal structure,
with the amplitude  varying over a wider interval.
It might therefore be tempting to expect that in the region of their 
coexistence ($h< \frac13$), the N\'eel wall should be
 unstable against the decay into the
``smoother" (Bloch) wall, and this is indeed the case in the
Ginzburg-Landau and Klein-Gordon counterparts of eq.(\ref{NLS})
\cite{Coullet,Hawrylak,Ivanov}. Surprisingly,
 the NLS dynamics turn out to be  very different.

We will show, numerically, that when $\gamma=0$,
 both walls can stably move with nonzero velocities and form stable
stationary and oscillatory, quiescent and travelling  bound states. 
The resulting bifurcation diagram will then be interpreted
using two integrals of motion of eq.(\ref{NLS}),
and a relation of our model to biaxial ferromagnets established.
Turning to the dissipative case, we will
 give an analytical proof of the stability of 
the N\'eel wall for all $h$ and $\gamma$ and 
describe stable bound states formed by the
damped  walls. $\Box$

\begin{figure}
\includegraphics[ height = 2in, width = 1.\linewidth]{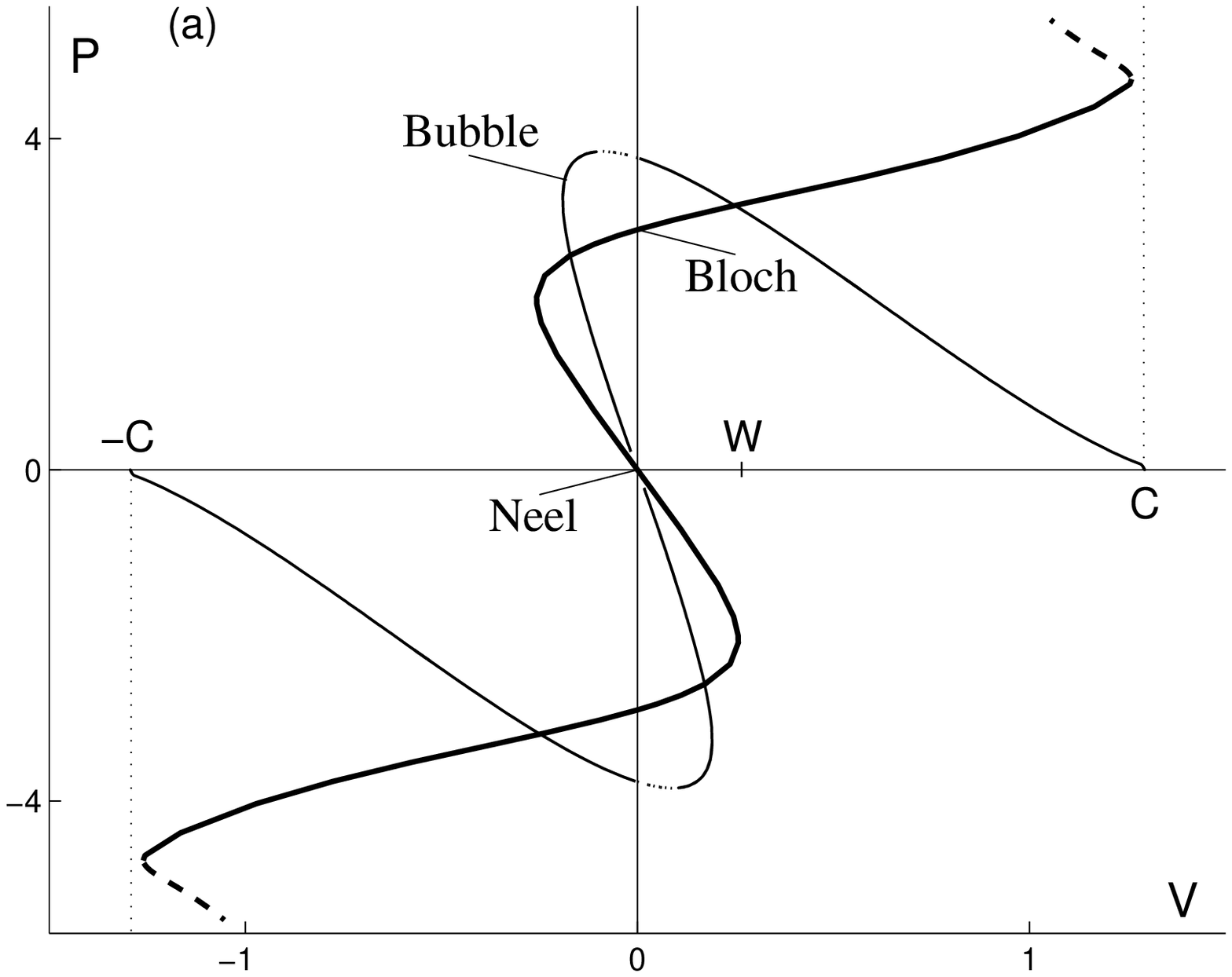}
\includegraphics[ height = 2in, width = 1.\linewidth]{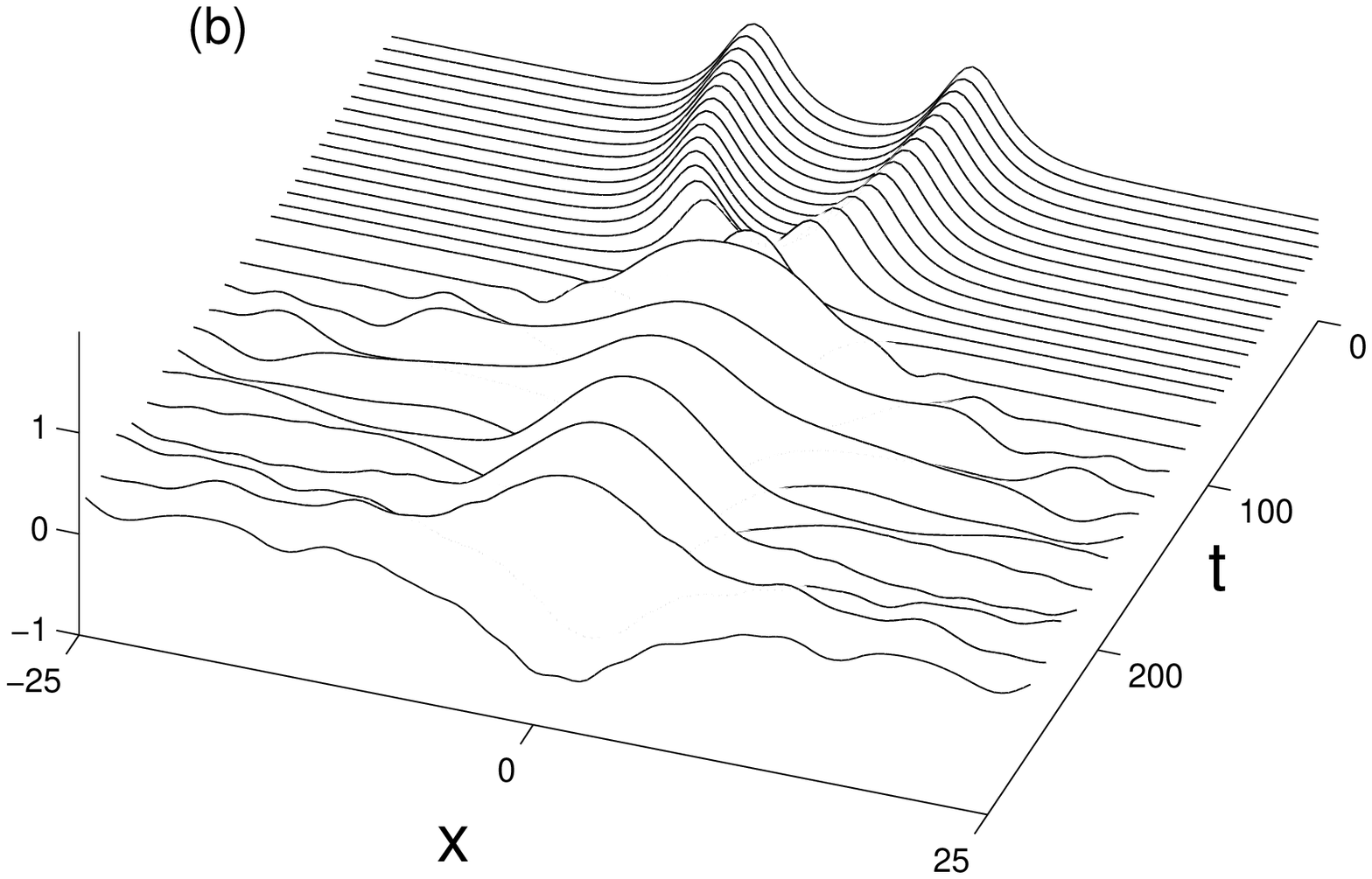}
\caption{\sf (a) The momentum of the travelling Bloch and N\'eel walls
(thick)
and  their nonoscillatory bubble-like complex (thin line).
For $|V|$ close to 
$c$, the wall
attaches  a small-amplitude bubble
on each flank; this accounts for the turn of the thick curve
near $|V|=c$. The dotted segments of the continuous branches indicate 
unstable solutions. (b) The formation of an oscillatory
breather-like complex
of two walls. (Only the real part of $\psi$ is shown for visual
clarity.) In (a), $h=\frac{1}{15}$; in (b), $h=0.1$.
}
\label{fig1}
\end{figure}

Both static kinks, (\ref{3}) and (\ref{19}), belong to a broader
class of uniformly moving solutions of the form $\psi(x-Vt)$. 
We found these by 
solving  equation
$\frac12 \psi_{xx}-iV\psi_x-|\psi|^2\psi+\psi=h\psi^*$
\cite{numerics}. 
Fig.\ref{fig1}(a) shows the  momentum of the 
travelling wall as a function of its velocity. The momentum
$P =
{\rm Im\/} \int {\psi_x}  \psi^* dx$ is one 
of the two conserved quantities of eq.(\ref{3}) with $\gamma=0$, and hence
is a natural choice for the integral characteristic of its solutions.
The second integral is
energy, and it will also be used below:
\begin{equation}
 E = {\rm Re\/} \int \left( \frac{|\psi_x|^2}{2} + \frac{|\psi|^4}{2} -
|\psi|^2 + h \psi^2 + \frac{A^4}{2} \right)dx.
\label{E}
\end{equation}
(Here $A^2=1+h$.)
The
stability of the travelling walls was examined \cite{numerics}
 by computing eigenvalues $\lambda$ of
\begin{equation}
 \mathcal{H} \vec{\varphi} = \lambda J \vec{\varphi}, 
  \label{starstar} 
\end{equation}
where   the column $\vec{\varphi} = (u,v)^{T}$, 
 the operator ${\cal H}$ is given by
\[
{\cal H}= - \frac{I}{2}  \partial_x^2 
+ \left(
\begin{array}{lr}
   3 \mathcal{R}^2 + \mathcal{I}^2+h &
 2 \mathcal{RI}-V \partial_x +\gamma \\  2 \mathcal{RI} + V \partial_x
 -\gamma  &
  \mathcal{R}^2 + 3\mathcal{I}^2-h
\end{array}
\right),
\]
and
$J$ is the
antisymmetric matrix with $J_{21}=-J_{12}= 1$.
Eq.(\ref{starstar}) is obtained by linearising eq.(\ref{NLS})  about
$\psi =
\mathcal{R} + i \mathcal{I}$ in the co-moving frame, and letting
$\delta\psi = (u+iv)e^{\lambda t}$.

In a striking contrast to the diffusive
and relativistic dynamics \cite{Hawrylak,Coullet,Ivanov}, our
numerical analysis  of eq.(\ref{starstar}) reveals that not only
the stationary N\'eel wall, but  the 
entire  branch
of travelling
kinks in Fig.\ref{fig1}(a)
is stable. 
This multistability admits a simple explanation in terms of
the energy and momentum, though.
The energy of the stationary N\'eel wall, $E_N =
\frac{4}{3}(1+h)^{3/2}$,
 is greater than
that of the stationary Bloch wall, $E_B =
4\sqrt{h} - \frac{4}{3}h^{3/2}$, and so  one might expect
$\psi_N$ to decay into $\psi_B$ plus radiation waves ---
as in the relativistic case \cite{Hawrylak}. However, unlike their
relativistic counterparts, our Bloch and N\'eel walls have unequal
momenta, with $P_B > P_N$  (see Fig.\ref{fig1}(a)) --- and this
makes the $\psi_N \to \psi_B$ decay impossible.

Next, our  simulations of the time-dependent
eq.(\ref{NLS})  show that 
two stationary
Bloch walls with opposite chiralities
(i.e. opposite signs in (\ref{19})) can
attract and form a motionless breather-like
bound state
(Fig.\ref{fig1}(b)).
An attraction of  stationary Bloch and  N\'eel walls 
 results in a {\it moving\/} breather. 
There also exist nonoscillatory, bubble-like, bound states
of $\psi_N$ and $\psi_B$. Note that unlike their parent walls,
all of  these complexes approach the
{\it same\/}
background
  as $x\rightarrow  \infty$ and $x\rightarrow  -\infty$. 
The bubble-like solitons admit most
transparent physical interpretation: they describe ``islands" of one
stable phase in the sea of the other one, e.g. patches oscillating 
$180^\circ$ out of phase with the rest of
the vibrated water channel or chain of
pendula.  
Below we focus on the bubbles and  
relegate the breathers to a separate publication.

  For each $h$
there is a one-parameter family of motionless
bubbles, the parameter being the separation distance $z$ between the two
walls. (Accordingly, there are two zero eigenvalues in the spectrum of the
operator (\ref{starstar})
associated with each bubble, one translational and the other one
corresponding to variations in $z$). There is a particular separation $z =
\zeta$ for which the bubble is symmetric: $\psi_{\zeta}^*(-x) =
-\psi_{\zeta}(x)$. 
The symmetric bubble turns out to have the largest momentum over bubbles with
various $z$, while $\zeta$ is the smallest possible separation: $z \geq
\zeta$. More importantly, 
 it is the  only {\it stable\/} bubble. All nonsymmetric bubbles ($z >
\zeta$) were found to have a pair of nonzero real eigenvalues $\pm\lambda$
in their spectrum. As $z \rightarrow \zeta$, the pair converges at the origin and
so the symmetric bubble has {\it four\/} zero eigenvalues.

Let $\pm \lambda$ be a pair of eigenvalues diverging from zero as $z$ grows
from $\zeta$ and
 assume  that $\lambda = \epsilon^{1/2} \lambda_1 + \dots$
for small $\epsilon = z - \zeta$. Then the associated
eigenfunction ${\vec \varphi}$ expands as ${\vec \varphi} = {\vec \varphi}_0 +
\epsilon^{1/2}{\vec \varphi}_1 + \dots$, where 
${\vec \varphi}_0$ is a linear combination of the two zero modes:
${\vec \varphi}_0 = (C_1
\partial_x {\vec \psi} + C_2 \partial_z {\vec \psi})|_{z = \zeta}$,
with ${\vec \psi} \equiv
 ({\cal R},{\cal I})^T$. (The other two zero eigenvalues  have only
 {\it generalised\/} eigenvectors associated with them.)
Substituting this into (\ref{starstar})
and using  $\mathcal{H}_z =
\mathcal{H}_{\zeta} + \epsilon\mathcal{H}_1 + \dots$,
the order $\epsilon^{1/2}$ yields
$\mathcal{H}_{\zeta} {\vec \varphi}_1 = \lambda_1 J {\vec \varphi}_0$. 
A bounded ${\vec \varphi}_1(x)$ exists only 
if $\lambda_1 J {\vec \varphi}_0$ is orthogonal both
to
  $ \partial_z {\vec \psi}$ and 
$\partial_x {\vec \psi}$.
This orthogonality condition amounts to $(dP/dz)|_{z = \zeta} =
0$, 
and the latter relation explains why the momentum
has to reach its maximum at the value of $z$ for which 
a pair of real eigenvalues converges at the origin.

The other implication of the relation $dP/dz = 0$  is that
it allows the
symmetric bubble to be continued to $V \neq 0$ \cite{Baer}. 
The resulting branch
of moving bubbles is shown in Fig.\ref{fig1}(a).
 As $|V| \rightarrow c =({1+2h+ \sqrt{4h(1+h)}})^{1/2}$,
which is the minimum phase velocity of linear waves,
the bubble
degenerates into the flat background,
whereas when $V,P \to 0$, it transforms into a  pair of N\'eel walls
with the separation $z \to \infty$. The entire branch of moving
bubbles is stable, with the exception of a small region
 between $V=0$ and the point of the maximum
$|P|$ inside which a  real pair $\pm \lambda$ occurs
(Fig.\ref{fig1}(a)).
The change of stability
 at  points where  $dP/dV=0$, is explained
  in \cite{Baer}.
  
 The diagram Fig.\ref{fig1}(a) can further justify
referring to the kinks (\ref{3}) and (\ref{19}) as N\'eel and Bloch walls.
In fact stationary nonchiral interfaces called N\'eel walls are known in 
uniaxial ferromagnets, where they coexist with chiral (Bloch) walls.
When the axial symmetry is broken, the two types of walls can move;
this occurs 
in particular 
in easy-axis 
 ferromagnets with the second, weaker, anisotropy axis
 ($\beta,\epsilon<0$ and $H=0$ in eq.(\ref{LLE2}) below) \cite{KIK}.
 The $P(V)$ curve for the easy-axis walls is qualitatively similar 
 to our  Fig.\ref{fig1}(a):
 The N\'eel wall's momentum $P_N= 0$ while $P_B \neq 0$;
as the velocity
$V$ grows, the two branches are drawn closer
together and  finally  merge. The limit velocity $V=w$ is known as the Walker's
velocity \cite{KIK}. The $E(V)$ curves are also similar.

 This analogy suggests that there could be a link
 between the time-dependent NLS (\ref{3}) and Heisenberg
 ferromagnets
 and indeed, there is one.
Consider
  a  quasi-one-dimensional ferromagnet 
with a weakly anisotropic easy plane $(M_x,M_y)$, in
 the external stationary magnetic field  along $M_z$. The magnetisation
 vector ${\bf M}=(M_x,M_y,M_z)$ lies on the sphere, ${\bf M}^2=M_0^2$,
 and satisfies the (damped) Landau-Lifshitz equation \cite{KIK}:
 \begin{eqnarray}
 \frac{\hbar }{2 \mu_0} {\bf M}_\tau=
 {\bf M} \times \frac{\delta }{\delta {\bf M}}
 \int {\cal W} \ d \xi- \lambda 
 {\bf M} \times {\bf M}_\tau, 
 \label{LLE1} \\
 {\cal W}= \frac{\alpha}{2} (\partial_\xi {\bf M})^2+ \frac{\beta}{2} M_z^2 +
 \frac{ \epsilon \beta}{2} M_x^2 -HM_z +{\cal W}_0,
 \label{LLE2}
 \end{eqnarray}
 with $\beta>0$. 
If the anisotropy parameter $\epsilon$ is small and
 the  field $H$ is close to  $\beta M_0$: 
$H= \beta M_0-\epsilon q$, the vector ${\bf M}$ will stay near
  the northern pole of the sphere. Choosing
 $s \equiv qM_0-\beta M_0^2/2>0$ for $\epsilon>0$ and $s<0$
 for $\epsilon<0$,
we define $M_x+iM_y= (2\epsilon s/\beta)^{1/2}  \psi^*$.
Assuming that
the relaxation constant $\lambda$ is ${\cal O}(\epsilon^{1/2})$
or smaller, and that ${\bf M}$  depends only on ``slow"
variables $x=(\epsilon s/2 \alpha M_0^2)^{1/2} \xi$ and 
$t=(2\epsilon  \mu_0 s/ \hbar M_0) \tau$, 
 eq.(\ref{LLE1})-(\ref{LLE2}) reduces to eq.(\ref{NLS})
with   $h=\beta M_0^2/(2s)$ and $\gamma=0$.
Note that the resulting NLS  is undamped --- 
although the original Landau-Lifshitz equation did
include a small damping term. The effect of
damping will become noticeable only on  time scales longer than
$\epsilon^{-1}$; these are not captured by eq.(\ref{NLS}).
Also note that despite the
 analogy between the easy-axis and easy-plane
 ferromagnets, there are important physical differences.
 In particular  in the easy-axis case the walls interpolate 
 between  ${\bf M}/M_0= \pm (0,0,1)$ while in our case they separate domains 
 with ${\bf M} \sim (0, \pm |\epsilon|^{1/2},1)$. $\Box$
 
Proceeding to the damped situation, $\gamma \neq 0$, 
our first goal  is to demonstrate the
stability of  the N\'eel wall.
We let $\psi(x,t) = \psi_N(x) +
\delta\psi(x,t)$, where
\begin{equation}
\delta\psi(x,t) = [u(X) + i v(X)] e^{(\mu-\Gamma)T - i\theta};
\label{4}
\end{equation}
$X = Ax$, $T = A^2t$, $\Gamma = A^{-2} \gamma$ and $\mu$ is complex.
Linearising eq.(\ref{NLS}) in small $\delta\psi$\space we obtain an eigenvalue problem
\begin{equation}
(L_0 + \epsilon)v = (\mu - \Gamma)u, \quad L_1 u = -(\mu+\Gamma)v,
\label{5}
\end{equation}
where $\epsilon = 2 - 2/A^2$\space and $L_0$\space and $L_1$\space are the
Schr\"{o}dinger operators with familiar spectral properties:
\begin{eqnarray*}
 L_0 \equiv - {\textstyle \frac12} 
 \partial_X^2 -  \mbox{sech}^2{X}, \quad
 L_1 \equiv L_0+ 2 \, \mbox{tanh}^2{X}.
\end{eqnarray*}
Introducing $\nu^2 = \mu^2 - \Gamma^2$\space and $w = \nu^{-1}(\mu +
\Gamma)v$ \cite{BBK}, we eliminate $\Gamma$\space from the eigenvalue problem
(\ref{5}):
\begin{equation}
(L_0 + \epsilon)w = \nu u, \quad L_1 u = -\nu w.
\label{8}
\end{equation}
Now we will show that $\nu^2 <0$\space for all $0 \leq \epsilon < 2$,
so that $\mu^2 < \Gamma^2$\space and all perturbations decay to zero as $t
\rightarrow \infty$.

The operator $L_1$\space has a zero eigenvalue, with the associated
eigenfunction $y_0(X) = \mbox{sech}^2 X$, and no negative eigenvalues.
Consequently, on the subspace $\mathcal{R}$\space defined by 
\begin{equation}
\int u(X) y_0(X) dX = 0,
\label{9}
\end{equation}
there exists an inverse operator $L_1^{-1}$\space and so (\ref{8}) becomes
$(L_0 + \epsilon) u = -\nu^2 L_1^{-1} u$,
with $L_0 + \epsilon$\space symmetric and $L_1^{-1}$ a positive operator.
The smallest eigenvalue $-\nu_0^2$ is given by the minimum of the Rayleigh
quotient:
\begin{equation}
-\nu_0^{2} = \min_{u\in\mathcal{R}}\frac{\int u(L_0 + \epsilon)u dX}{\int
u L_1^{-1}u dX}.
\label{11}
\end{equation}
To prove that $-\nu_0^2 > 0$ it is sufficient to show that the minimum of the
quadratic form $\int{u(L_0 + \epsilon)u dX}$ is positive on $\mathcal{R}$
 \cite{Vahitov}. Assuming that $u(X)$ are normalised
 by $\int u^2 dX = 1$,
the minimum
is attained on the solution $u(X)$ to the nonhomogeneous boundary-value
problem
\begin{equation}
(L_0 + \epsilon) u(X) = \eta u(X) + \alpha y_0(X),
\label{12}
\end{equation}
where $\eta$ and $\alpha$ are the Lagrange multipliers. The minimum
equals $\eta$ --- provided $\eta$ and $\alpha$ are chosen so that the
$u(X)$ satisfies eq.(\ref{9}) and the normalization constraint.

The operator $L_0$ has a single discrete eigenvalue $E_0 =
-\frac{1}{2}$ with the eigenfunction $z_0(X) = (1/\sqrt{2})\mbox{sech}X$,
and the continuous spectrum of eigenvalues $E(k) = k^2$, with
 \begin{equation}
z_k(X) = \frac{ik + \tanh X}{ik - 1} e^{-i k x}, \quad -\infty < k < \infty.
\label{13}
\end{equation}
Expanding $y_0$ and $u$ over the complete set $\{z_0;
z_k\}$\space gives
\begin{eqnarray*}
 y_0(X) = Y_0 z_0(X) + {\int } Y(k) z_k(X) dk, \\
u(X) = U_0 z_0(X) + { \int } U(k) z_k(X) dk.
\end{eqnarray*}
Substituting into (\ref{12}) and using the orthogonality of the
functions in the set produces $U(k) = \alpha (k^2 + \epsilon - \eta)^{-1}Y(k)$
and $U_0 = \alpha (E_0 + \epsilon - \eta)^{-1}Y_0$. Using these in (\ref{9})
gives
\begin{equation}
{\mathfrak g}_{\epsilon}(\eta) \equiv \frac{Y_0^2}{E_0 + \epsilon - \eta} +
\int^{\infty}_{-\infty} \frac{\abs{Y(k)}^2}{k^2 + \epsilon - \eta}dk = 0.
\label{14}
\end{equation}

The minimum of the quadratic form $\int u (L_0 + \epsilon) u dX$ is given
by the smallest root $\eta^{*}$ of the function (\ref{14}). The function
${\mathfrak g}_{\epsilon}(\eta)$ is increasing for $-\infty < \eta \leq
\epsilon$, apart from the point $\eta = E_0 + \epsilon$ where it drops
from $+\infty$ to $-\infty$. As $\eta \rightarrow -\infty$,
${\mathfrak g}_{\epsilon}(\eta) \rightarrow +0$; as $\eta \rightarrow \epsilon$,
${\mathfrak g}_{\epsilon}(\eta)$ tends to a finite value. (This follows from
the fact that
\[ Y(k) = \frac{ik}{1 + ik} \, \frac{\pi k /2}{{\rm sinh}(\pi k/2)}, \]
hence the integral in (\ref{14}) converges for all $\eta \leq \epsilon$.)
Consequently, there is only one root $\eta^{*}$ and its sign is opposite
to the sign of ${\mathfrak g}_{\epsilon} (0)$. Since $\partial {\mathfrak
g}_{\epsilon}(\eta)/\partial\epsilon < 0$, we have ${\mathfrak g}_{\epsilon} (0) <
{\mathfrak g}_0 (0)$ while the value ${\mathfrak g}_0 (0)$ can be
calculated as
\begin{equation}
{\mathfrak g}_0 (0) = \frac{Y_0^2}{E_0} + \int^{\infty}_{-\infty}
\frac{\abs{Y(k)}^2}{k^2}dk = \int y_0 L_0^{-1}y_0 dX.
\label{15}
\end{equation}
Noticing that $L_0^{-1}y_0(X) = -1 + c \tanh{X}$, with $c$
an arbitrary constant, eq.(\ref{15}) yields ${\mathfrak g}_0 (0) = -2$ and hence
$\eta^{*}$ cannot be negative for any $\epsilon$. Thus $-\nu^2 > 0$
and the N\'eel wall is stable for all $h$ and $\gamma$
(with $h \ge \gamma \ge 0$).

Are there any other attractors for nonzero $\gamma$?
When $\gamma \neq 0$, the momentum is, in general, changing with
time: $\dot{P} = -2\gamma P$, and therefore a uniformly moving soliton
  has to satisfy $P = 0$. Since both  curves in
Fig.\ref{fig1}(a) cross the $P = 0$ axis only at the
stationary N\'eel walls, we conclude that
no other solutions persist for small nonzero $\gamma$.
This  does not, however, exclude the existence of new solutions
for {\it larger\/} $\gamma$.
Our numerical analysis has, in fact, revealed a window of
$h$ values, $\gamma < h \, {^< \!\!  \!\!_\sim} \, 0.35 +0.8 \gamma$
(with $0.1 \le \gamma \le 0.85$) where
two N\'eel walls attract and form a stable stationary bubble.
$\Box$


In conclusion, the remarkable stability  
of the damped-driven kinks and their
bound states is in sharp contrast  with stability
properties of the  bright solitons. 
The stable coexistence of two types of 
 domain walls and their  complexes in the 
undamped case is also worth emphasising.
This multistability is not observed in the 
parametrically driven  
Klein-Gordon and Ginzburg-Landau equations
and is due to the availability of the 
momentum integral which takes different values on different solutions.

We thank Nora Alexeeva for writing 
a pseudospectral code for the time-dependent NLS  (\ref{NLS})
and Boris Ivanov for  useful comments on this work.
IB  was supported by the NRF 
and URC grants; 
SW by the  NRF and the
Joseph Stone
fellowship; EZ by the RFFI grant 000100617.


\begin{thebibliography}{99}
\bibitem{numerics}
In solutions of the stationary eq.(\ref{3})
we used a  fourth-order Newtonian iteration, on the interval
(-100,100), with the stepsize $\Delta x=5 \cdot
10^{-3}$.
The eigenvalue problem (\ref{starstar}) was solved by the Fourier expansion
over 600 modes.
\end{thebibliography}
\end{document}